\documentclass[draftcls,onecolumn,12pt]{IEEEtran}

\usepackage{caption}
\usepackage{amssymb,amsmath,epsfig,graphics,graphicx,enumerate,float,subcaption,verbatim,xspace, color, hyperref,url}
\usepackage{epstopdf}

\newcommand{\nop}[1]{}

\title{Detecting Clusters of Anomalies on Low-Dimensional Feature Subsets with Application
to Network Traffic Flow Data\thanks{This work supported
in part by a Cisco Systems URP gift.}}
\author{Zhicong Qiu, David J. Miller and George Kesidis\\
School of EECS, The Pennsylvania State University\\ 
\{zzq101,djm25,gik2\}@psu.edu}

\begin{document}         
\maketitle
\begin{abstract}
In a variety of applications, one desires to detect groups of anomalous data samples,
with a group potentially manifesting its atypicality (relative to a reference model) on a low-dimensional subset
of the full measured set of features.  Samples may only be weakly atypical individually,
whereas they may be strongly atypical when considered jointly.
What makes this group anomaly detection problem quite challenging is that 
it is {\it a priori} unknown which subset of features jointly manifests a particular 
group of anomalies.  Moreover, it is unknown how many anomalous groups are present in a given data batch.
In this work,
we develop a group anomaly detection (GAD) scheme to identify the subset of samples and subset of features that jointly specify an anomalous cluster.  We apply our approach to 
network intrusion detection to detect BotNet and peer-to-peer flow clusters.
Unlike previous studies, our approach captures and exploits statistical dependencies
that may exist between the measured features.
Experiments on real world network traffic data demonstrate the advantage of our proposed system, and 
highlight the importance of exploiting feature dependency structure, compared to the feature (or test) independence assumption made in previous studies.

\end{abstract}


\section{Introduction}
Group anomaly detection has recently attracted much attention, with applications in astronomy \cite{Hierarchical},
social media \cite{GLAD}, disease/custom control \cite{CMU}\cite{DAS} and network intrusion detection \cite{Miller}\cite{Fatih}\cite{Eskin2000anomaly}.
In this work, we focus on group anomaly detection applied to network intrusion detection, 
where the anomalous groups are either distributed Botnet (Zeus) or peer-to-peer (P2P) nodes generating traffic that deviates from the normal (Web traffic) behavior.
%It is a challenging problem because these malicious nodes will hide themselves as much as possible among the normal ones, hence, they tend to be slightly anomalous when investigated individually. 
%Moreover, unlike the case in \cite{OC-MM}, it is a priori unknown how many anomalous groups are present in the given domain and which subset of features and nodes generate anomalous traffic as a group.
Many existing intrusion detection systems (IDSs) only make sample-wise anomaly detections, e.g., in \cite{Paxson}, the samples which deviate most from a normal (reference) model are flagged as anomalies/outliers.
However, such an approach does not identify anomalous {\it groups} ({\it e.g.}, a collection
of BotNet flows), whose samples all exhibit similar behavior.  Identifying such groups
could be essential for mounting some form of system response or defense. 
Moreover, 
individual samples may only be weakly atypical.  Thus, a sample-wise IDS may either
fail to detect most of the anomalous samples, or may incur high false positives
when a low detection threshold is used. 
By contrast, (weakly) anomalous samples whose anomalies are all ``similar to each other'' 
may be {\it strongly} atypical when considered in aggregate, {\it i.e.} jointly.
For example, for an $N=100$-dimensional feature space, 
suppose there is a sizeable collection of samples in the captured data batch that are all (even only weakly) atypical with respect to the
{\it same} feature or the same (small) feature subset.  There is a low probability that this occurs by chance, ({\it i.e.},under the null).  Thus, such {\it clusters of anomalies}, each defined by a sample subset
and a feature subset, may be strongly atypical, and hence more convincing anomalies, than
individual sample anomalies.
It should be noted that there is an enormous number of candidate anomalous clusters, considering the
conjoining of all possible sample subsets and all possible feature subsets.
Thus, a GAD scheme will require some type of heuristic search over this huge space,
aiming to detect the most statistically significant cluster candidates. 
In the sequel, we propose such a GAD scheme.
Rather than assuming individual features or outlier events
are statistically independent under the null
as in \cite{CMU,Fatih}, in our approach, as in \cite{provisional}, 
we capture and exploit
statistical dependencies amongst the features defining a candidate cluster.
Compared to previous works, as shown in our experiments, 
the proposed scheme is more effective in detecting group anomalies.

The paper is organized as follows.  Section II defines the problem and elaborates on related works. 
Section III describes the proposed model. 
Section IV evaluates the system performance, and compares with some recent works. 
We then discuss some extensions of our system and future works in section V, followed by conclusions.
\section{Problem Definition and Related Work}
We assume there is a batch of normal web traffic available at the outset as training set,
{\it i.e.}  $X_l = \{\underline{\tilde{x}}_i, i=1,...,T_l, \underline{\tilde{x}}_i\in R^D\}$, where $\underline{\tilde{x}}_i$
is a $D$-dimensional feature vector representing the $i$-th training traffic flow\footnote{A flow is a bidirectional communication sequence
between a pair of nodes in a network.}, and where we assume the number of training flows $T_l$ is large enough to learn an accurate reference model (null hypothesis).
These traffic flows can either be generated and captured in a sandbox environment, or sampled from a domain of interest (data warehouse, enterprise network) in real time under normal operating conditions.
Given a model of normal network traffic learned based on ${\cal X}_l$, our goal is to interrogate 
a capture batch of unknown traffic flows ${\cal X}_u= \{\underline{x}_i, i=1,...,T_u, \underline{x}_i\in R^D\}$\footnote{Unknown in the sense that we do not know
which if any of these flows represent outliers or attacks.},
seeking to identify latent groups of Botnet or P2P traffic, with the flows in each such group exhibiting similar behavior.
This has been previously considered in \cite{Fatih}, where the authors used the samples in ${\cal X}_l$ to estimate bivariate Gaussian Mixture Models (GMMs), on all feature pairs, representing the null hypothesis.
These bivariate GMMs were used to evaluate {\it mixture-based p-values}\footnote{A p-value is the probability that an event is more extreme than the given observation.} for all pairs of features.
Assuming the features (tests) are statistically independent, a joint significance score function was defined for a given candidate cluster, specified by
its sample subset and feature subset, with a Bonferroni correction used to account for multiple testing. 
Instead of exhaustively searching over feature subset candidates at order $K$ \footnote{We use ``order'' to denote the maximum feature dimension considered.}, 
the authors proposed to trial-add individual features only to the {\it top-ranking} candidate feature subsets (in terms of the Bonferroni corrected score) at order $K-1$. 
Furthermore, the authors showed that the computational complexity of determining the optimal (in terms of the joint score) {\it sample} subset given the
feature subset fixed is linear in $T_u$, once the samples in a given feature subset are ranked by their aggregate p-values.
However, the independent test assumption used in \cite{Fatih} becomes grossly invalid as more and more features are included in 
a cluster, which limits the proposed model's detection accuracy for increasing $K$.
A related framework was also proposed in \cite{CMU}, albeit assuming categorical attributes. Here, the authors built a single, global null hypothesis Bayesian network based on ${\cal X}_l$.  They then assigned categorical-based p-values to samples in $X_u$, 
with a cross entropy based scoring criterion used to efficiently search for the best feature and sample subset candidates. 
A limitation of this approach is that the statistical tests are again assumed to be independent.

We herein describe and experiment with a method of anomaly detection
 that extends \cite{CMU,Fatih} and is closely related to 
\cite{provisional}. The method
captures dependencies between the features in a candidate cluster by a 
dependence tree structure, and uses this model to help evaluate joint p-values for cluster candidates. 
As in \cite{Fatih}, the Bonferroni corrected score is used as the objective function for evaluating the best cluster candidates (defined by their sample and feature subsets).  The candidate with the best such score is detected as a cluster of anomalies. 
Whereas in \cite{CMU} a single global (null model) Bayesian network is used to assess candidate clusters, in \cite{provisional} and in
the current work a local, customized {\em cluster-specific} dependence tree model is used to assess each candidate cluster.
     
\section{Proposed Model}
\subsection{Mixture-based P-values for Singletons and Feature Pairs}
Consider a (sample, feature) index pair $(i,j)$ and let $I_i^{(j)}$ be an indicator variable for the event that the $j^{th}$ feature value of the $i^{th}$ sample, $x_i^{(j)}$,
is an outlier with respect to the null distribution for feature $X^{(j)}$.
Let $O^{(j)}(x_i^{(j)})$ be a subset of the real line such that, $\forall y^{(j)} \in O^{(j)}(x_i^{(j)})$, $y^{(j)}$ is ``more extreme'' than the given observation $x_i^{(j)}$.
One good definition for this set, consistent with evaluating a 2-sided p-value for a unimodal, symmetric null for $X^{(j)}$, is:
\begin{eqnarray}
\label{2-sided}
O^{(j)}(x_i^{(j)};\mu^{(j)}) = \{y^{(j)}:|y^{(j)}-\mu^{(j)}|\geq |x_i^{(j)}-\mu^{(j)}|\}, \nonumber
\end{eqnarray}
where $\mu^{(j)}$ is a representative (mean) value for feature $X^{(j)}$. 
Given the component means $\mu_l^{(j)}, l=1,...,L_j$, of an $L_j$-component Gaussian mixture null, let $M^{(j)}(x)$ be a function that maps $x$ to the mixture component index set $\{1,2,...,L_j\}$,
i.e., it indicates which mixture component generated $x$.
Also, let $Y_j$ be a random variable distributed according to the mixture density $f_{X_j}(x)$.  
Then, for a given observation $x_i^{(j)}$, we define the binary random variable $I_i^{(j)}$, where
$I_i^{(j)}=1$ if $Y_j$ is more extreme under the null than $x_i^{(j)}$.  
Then, we can write
the 
singleton mixture p-value as: 
\begin{eqnarray}
\lefteqn{ P[I_i^{(j)}=1]} & & \nonumber \\
 &  = & P[Y_j \in \cup_{l=1}^L ((O^{(j)}(x_i^{(j)};\mu_l^{(j)})) \cap (M^{(j)}(x_i^{(j)})=l))] \nonumber
\\
& = & \sum\limits_{l=1}^L P[Y_j \in O^{(j)}(x_i^{(j)};\mu_l^{(j)})] P[M^{(j)}(x_i^{(j)})=l].  \label{mixture}
\end{eqnarray}
Here, an extreme outlier event is conditioned on $x_i^{(j)}$ having been generated by component density $l$.
The probability $P[Y_j \in O^{(j)}(x_i^{(j)};\mu_l^{(j)})]$ is the two-sided Gaussian p-value, integrating over the region
$|y-\mu_l^{(j)}|\geq |x_i^{(j)}-\mu_l^{(j)}|$, while $P[M^{(j)}(x_i^{(j)})=l]$ is the {\it a posteriori} probability
that $x_i^{(j)}$ was generated by component $l$.

\begin{figure}
\centering

\hspace{-4.5mm} \includegraphics[width=3in]{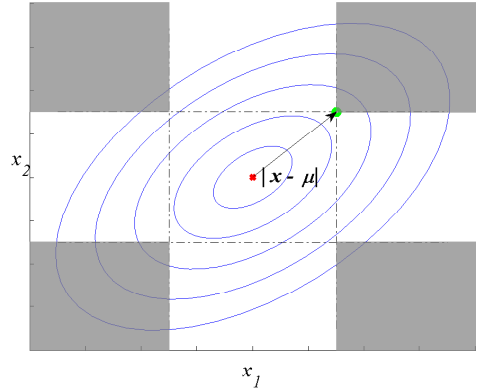}

\caption{Illustrative figure: bivariate Gaussian joint p-value measure coresponds to 
the four (unbounded) shaded corners in grey, with mean {\bf $\mu$} and a given observation {\bf $x$}}
\label{rect}
\end{figure}

Similarly, for a {\it pair} of observations $(x_i^{(j)},x_i^{(k)})$, we have the second order mixture p-value:
\begin{eqnarray*}
\lefteqn{P[I_i^{(j)}=1,I_i^{(k)}=1]~} & & \\
  & = & \sum\limits_{l=1}^L 
P[Y_j \in O^{(j)}(x_i^{(j)};\mu_l^{(j)}),Y_k \in O^{(k)}(x_i^{(k)};\mu_l^{(k)})]
\\
& & ~~~~~\cdot P[M^{(j,k)}(x_i^{(j)},x_i^{(k)})=l].
\end{eqnarray*}
Here, $P[Y_j \in O^{(j)}(x_i^{(j)};\mu_l^{(j)}),Y_k \in O^{(k)}(x_i^{(k)};\mu_l^{(k)})]$ integrates the $l$-th component bivariate Gaussian density over the
region 
\small
$$\{(y_j,y_k): |y_j - \mu_l^{(j)}| \geq |x_i^{(j)} - \mu_l^{(j)}|, |y_k - \mu_l^{(k)}| \geq |x_i^{(k)} - \mu_l^{(k)}|\}.$$
\normalsize
This region consists of the union of four unbounded rectangular regions in the plane, as illustrated in Figure \ref{rect}.

In this work, a sample's anomalousness on a given feature subset is estimated by a joint p-value, with statistical
dependencies between features accounted for by a dependence tree (DT) structure \cite{Chow}. 
Since the dependence tree \cite{Chow} is based on first and second order probabilities, the joint p-value will
be based on the singleton and second order mixture p-values, as given above.
A smaller joint p-value indicates a sample is more anomalous under the given feature subset.
\subsection{Scoring Clusters}
Let $\{I_c, J_c\}$ denote cluster candidate $c$, $I_c$ its sample subset and $J_c$ its feature subset. 
Let $T_c = |I_c|, N_c=|J_c|$. Note that p-values are uniformly distributed on $[0,1]$ under the null.
Thus, given a cluster with feature subset $J_c$, from a test batch of size $T_u$, the probability that at least one cluster with $T_c$ samples has a smaller p-value than $P[\underset{j\in J_c}{\cap}(I_i^{(j)})=1]$ is:
\begin{eqnarray}
\label{bonf}
1-(1-\prod_{i} P[\underset{j\in J_c}{\cap}(I_i^{(j)})=1])^{C(T_u,T_c)}
\end{eqnarray}
Here, $C(T_u,T_c) = \dbinom{T_u}{T_c}$, {\it i.e.} it is the number of combinations and implements multiple testing correction, accounting for all possible sample subset configurations in a cluster with $T_c$ samples, from a test batch of size $T_u$.
In principle, (\ref{bonf}) provides a sound basis at least for directly comparing all cluster candidates with the same feature subset $J_c$.
However, it does not allow comparing pairs of cluster candidates with any configurations of $(T_c, N_c)$, 
because all possible feature subset configurations at a given order, $N_c$,
have not yet been properly multiple-testing corrected.
Also, (\ref{bonf}) requires evaluation of the joint p-value $P[\underset{j\in J_c}{\cap}(I_i^{(j)})=1]), \forall i \in I_c$, 
which in general depends on the joint density function for $(X_{j_1},X_{j_2},...,X_{j_{N_c}}), j_m\in J_c, m=1,...,N_c$.
When $D$ is large, it is not practically feasible to learn and store these $\dbinom{D}{N_c}$ joint null density functions, 
i.e., for all possible combinations of features up to order $N_c$.
Thus, it appears some tractable representation of $P[\underset{j\in J_c}{\cap}(I_i^{(j)})=1])$ is needed. 
An obvious temptation is to assume that $I_i^{(j)}$ and $I_i^{(j')}$ are statistically independent $\forall j,j'\in J_c, j'\neq j$. 
But this is a very poor assumption, consistent with assuming the features are independent.

To address the above problems, we seek to modify (\ref{bonf}) in two respects.
First, we propose to multiple test correct both for the different sample and the different feature subsets, given a cluster candidate with $(T_c,N_c)$.
In this approach, instead of the exponent being the number of combinations, it becomes the product of combinations on samples and combinations on features.
Based on the Bonferroni approximation of (\ref{bonf}), we have {\it the joint score function} $S(I_c,J_c) = \dbinom{D}{N_c}\dbinom{T}{T_c}\prod_{i \in I_c}P[\underset{j\in J_c}{\cap}(I_i^{(j)})=1])$.  For 
this joint significance measure, we can efficiently determine the optimal {\it sample} subset, given a fixed feature subset, by greedy sequential sample inclusion, in sorted joint p-value order.
This is due to the unimodality of this Bonferroni approximated joint significance measure, as a function of the number of samples included in a cluster's sample subset (see next subsection). 

Second, a rich, tractable, joint probability mass function model that does capture statistical dependencies is a restricted form of Bayesian network, based exclusively on first and second order distributions,
i.e., the {\it dependence tree} (DT), which factorizes the joint distribution $P[\underset{j\in J_c}{\cap}(I_i^{(j)})=1])$ as a product of first and second order probabilities \cite{Chow}. 
In \cite{Chow}, it was shown that, even though there is an enormous number of unique dependence tree structures, one can efficiently find the globally optimal dependence tree, over all such structures,
maximizing the dataset's log-likelihood, by realizing that this can be recast as a maximum weight spanning tree problem, 
with the pairwise weights defined as the mutual information between the pairs of random variables.
The maximum weight spanning tree can be efficiently solved via Kruskal's algorithm, with complexity $O(N_c^2log(N_c))$. 
Hence, given any candidate feature subset $J_c$, Kruskal's algorithm can be applied to determine the DT that maximizes the likelihood measured on ${\cal X}_l$, 
i.e., the null hypothesis is determined, consistent with the given candidate feature subset $J_c$.

Based on a given DT structure,  $P[\underset{j\in J_c}{\cap}(I_i^{(j)})=1])$ factorizes as a product of first and second order distributions, i.e., $\forall i \in I_c$:
\begin{eqnarray}
\label{factorize}
P[\underset{j\in J_c}{\cap}(I_i^{(j)})] = P[I_i^{(j_1)}]P[I_i^{(j_2)}|I_i^{(j_1)}]...P[I_i^{(j_{N_c})}|I_i^{(j_{N_c-1})}],
\end{eqnarray}
where we use $j_1$ to denote the root node of the DT representing $J_c$.

It is apparent from (\ref{factorize}) that, for any feature subset, one can represent the joint p-value of a given sample by its first and second order mixture p-values.
That is, for any feature pair $(j,k)$, $P[I_i^{(j)}|I_i^{(k)}] = \frac{P[I_i^{(j)},I_i^{(k)}]}{P[I_i^{(k)}]}$.
The numerator and denominator are, respectively, the second and first order mixture-based p-values that we defined
earlier.  Also note that, in order to evaluate the first order mixture p-value $P[I_i^{(k)}]$,
we marginalize feature $j$ from the bivariate GMM for the feature pair $(j,k)$.  This gives us the GMM for feature $k$.
\subsection{Identifying the Optimal Sample Subset $I_c$, Given Fixed $J_c$}
Given a fixed $J_c$ and associated DT, we would like to choose the sample subset $I_c$ to minimize (\ref{bonf}). 
Applying the Bonferroni correction, this is essentially equivalent to choosing $I_c$
to minimize the joint score function: 
\begin{eqnarray}
\label{bonf_corr}
S(I_c,J_c)=\dbinom{D}{N_c}\dbinom{T_u}{T_c}\prod_{i \in I_c}P[\underset{j\in J_c}{\cap}(I_i^{(j)})=1]).
\end{eqnarray}

It is in fact easily shown that this objective function is globally minimized by the following procedure:
i) sort the samples in increasing order of their joint p-values $P[\underset{j\in J_c}{\cap}(I_i^{(j)})=1]$;
ii) sequentially include the samples on the sorted list into $I_c$, until the objective function no longer decreases. This procedure globally minimizes over $I_c$ given fixed $J_c$.
\subsection{Overall Search Algorithm}
First, using the normal samples in ${\cal X}_l$, all the first and second order null GMMs are separately trained
\footnote{Separately learning each marginal and pairwise feature GMM using the common training set  ${{\cal X}_l}$ will not ensure {\it consistency} with respect to feature marginalizations.  
Specifically, a {\it marginal-consistent} collection of univariate and bivariate density functions should satisfy the following: if we consider any feature pairs $(i,j)$ and ($j,k)$, marginalizing out
feature $i$ from the $(i,j)$ bivariate density and marginalizing out feature $k$ from the $(j,k)$ bivariate density should lead to the same marginal density for feature $j$.  
However, when the univariate and bivariate distributions are Gaussian mixtures, with a {\it non-convex} log-likelihood function (and with BIC-based model order selection separately applied to choose the number of components for each GMM), separate application of EM-plus-BIC to learn each GMM density function does not ensure a set of marginal-consistent distributions.  
This property is not centrally important here, however, since our main concern is only to learn marginal and pairwise density functions that allow accurate assessment of p-values.  
Accordingly, in this work we will apply EM-plus-BIC separately, to learn each low-order GMM.  

One approach to obtain marginal-consistent low-order distributions is to simply learn the single GMM for the joint distribution on the {\it full} feature vector, 
$\underline{X}$.  This {\it determines} (via marginalization) all lower-order distributions (which are also GMMs, and which are 
guaranteed to be marginal-consistent).  However, this strategy suffers from the curse of dimensionality.
Alternatively, we refer the interested reader to \cite{provisional}, where a procedure for {\it directly}, jointly learning a marginal-consistent set of low-order GMMs is elaborated. 
}.
Mutual information for all feature pairs is then calculated based on the bivariate GMMs.
This is achieved by generating $M=10^6$ samples from a given bivariate GMM distribution, and then estimating the mutual information
by $\frac{1}{M}\sum\limits_{n=1}^M \log(\frac{f_{X_1 X_2}(x_1^{(n)},x_2^{(n)})}{f_{X_1}(x_1^{(n)}) f_{X_2}(x_2^{(n)})}$. 
We then detect clusters in ${\cal X}_u$ sequentially, in a rank-prioritized fashion, according to the joint score $S(I_c,J_c)$. 
The algorithm operates on an enormous space of candidate clusters even if the feature space itself is only modestly sized $(D)$.
We start by sweeping over feature subset candidates at low orders and, for tractability, only the ``most promising'' candidates at higher orders, with candidate feature subsets at order $K$ formed by ``accreting''
new features to the best-scoring candidates at order $K-1$.
For each candidate feature subset $J_c$, its DT is first learned and its associated, optimal subset $I_c$ is then determined using
the method described in section III.C.
Evaluating all candidates at all feature subset orders, the one with the best score function value at each order $N_c$ is recorded.
The cluster with smallest Bonferroni-corrected score $S(I_c,J_c)$ is then forwarded as detected.
Its samples are then removed from the test batch.
Subsequent cluster detections can then be made following the same procedure.
Cluster detections are thus made (in general) in order of decreasing joint significance.

\section{Experimental Setup and Results}
Our experiments focus on detecting Zeus 
botnet and P2P traffic among normal Web traffic.
The Web packet-flows are 
obtained from the LBNL repository \cite{LBNL}. 
This dataset contains Web traffic on TCP port 80, 
with specified time-of-day information.
Specifically, the experiments in this paper are based on three datasets named 
``200412215-0510.port008'',
``20041215-1343.port008'' and ``20041215-1443.port010''.
The protocols to obtain normal, P2P and BotNet network traffic are the same as in \cite{Fatih},
i.e., we used the port-mapper in \cite{Zou2011} to identify P2P traffic in these files by a C4.5 decision tree pre-trained in another domain (the Cambridge dataset \cite{Li2009}).
The Zeus Botnet traffic are obtained from another domain \cite{Zeus2}.

\subsection{Feature Space Selection and Representation}

Firstly, we did {\it not} use layer-4 port number features for purposes of
detection \cite{Zou2011,salt}.
Also, we did not consider timing 
information herein because the Zeus activity
was recorded on another domain \cite{salt}.
In \cite{salt}, previous efforts were made to detect BotNet and P2P traffic using the well-known feature representation for network intrusion detection from \cite{efficient}.
The authors found that these features, though able to detect some
 attack activity, 
could not successfully discriminate BotNet or P2P from normal 
Web traffic, 
i.e., BotNet and P2P traffic appear as ``normal" 
Web activity according to the 
features of \cite{efficient,salt}.

To capture the intrinsic behavior of BotNet and P2P 
packet-traffic, we note that most Zeus BotNet traffic involves masters giving command (control) messages, while slaves execute the given commands.
In the case of P2P, nodes often
 communicate in a bidirectional manner, exchanging relatively large packets 
in both directions. 
Normal/background Web traffic, on the other hand, tends to involve
 server-to-client communications.

Hence, we seek to preserve the bidirectional packet size sequence information as feature representation for different traffic flows.
This feature representation was previously considered in \cite{Fatih,Miller}.
The authors used the first $N$ (we set $N=10$ in our experiments) packets after the three-way hand shake of each TCP flow.
Then a feature vector of dimension $2N$ is defined, specified by the sizes and directionalities of these $N$ packets.
Traffic are assumed to be alternating between client-to-server (CS) and server-to-client (SC).
A zero packet size is thus inserted between two consecutive packets in the same direction to indicate an absence of a packet in the other direction.
For example, if the bidirectional traffic is strictly SC, a zero will be inserted after each SC packet size.
This $2N$-dimensional feature representation preserves bidirectional information of a given TCP flow, which is essential for 
discriminating between P2P, Zeus and normal Web traffic.

\subsection{Performance Metrics}
Our algorithm detects clusters (groups) in a sequential fashion. 
For each extracted group, we rank the samples in the group by their associated joint p-values on the given feature subset.
These samples will be sequentially removed from the test batch, with the system then continuing to extract groups until the test set is depleted.
Then we sweep out an ROC curve based on these rank-ordered detected samples. 
A larger area under the ROC curve indicates earlier detections of anomalous groups, which implies the effectiveness of the intrusion detection system.
We compare our system's performance with a GMM based anomaly detector, trained by normal samples, on the whole feature space.
For this detector, we rank the test samples based on their data likelihood under the GMM, and sweep out an ROC curve.
We also compare with the approach presented in \cite{Fatih}, which assumes significance tests are independent (denoted ``Independence tests''), 
and with the recent work presented in \cite{CMU} with a slight modification --
instead of discretizing feature values consistent with \cite{CMU}, we use a single dependence tree null distribution learned on 
${\cal X}_l$ and our proposed joint p-value for continuous features, $P[\underset{j\in J_c}{\cap}(I_i^{(j)})=1]$.
We denote this variation on the approach in \cite{CMU} by ``single Bayesian Net.''
There are two generalization performance measures of interest on the test set:
one is the aforementioned ROC area under curve (ROC AUC) as a function of the maximum feature subset size for a cluster, $K_{\rm max}$.
The other is the top 100 precision rate, defined as the fraction of anomalous samples amongst the first 100 detected samples. 
Lastly, instead of exhaustively searching over all feature subsets at order $K$, we trial-add individual features to the top candidate feature subsets from order $K-1$.
At each order $K$, starting from order 2, we only consider the top 500 candidates from order $K-1$.

Two different sets of experiments were performed, one on synthetic data, and the other on the network data mentioned earlier.
In the synthetic dataset experiment, we used one unimodal Gaussian with 10 dimensions to generate normal samples 
and two additional unimodal Gaussians to generate two distinct anomalous clusters.  The two anomalous clusters use the same
distribution as the normal distribution for nine of the ten features.  Thus, they deviate from the normal (null) distribution
only on a single feature dimension (this ``informative'' feature dimension was different for the two clusters). 
Their corresponding sample subsets consist of 2.5\% of the whole data batch ${\cal X}_u$ (so the proportion of anomalous samples in ${\cal X}_u$ is 5\% of the total). 
The variance of the informative features was chosen to be the same as that of the normal features, $\sigma_n^2$.
Moreover the mean of the informative feature under an anomalous cluster was chosen to be two standard deviations 
away from the mean under the normal class, {\it i.e.} $|\mu_n - \mu_a| = 2 \sigma_n$,
where we use subscripts $n$ and $a$ to denote `normal' and `anomalous', respectively.
Thus, if we consider only the informative feature dimension, the Bayes error rate in discriminating normal from anomalous
is 15.87\%.
After generating the synthetic data batch (with a size of ten thousand samples), we randomly chose 20\% of normal samples as ground-truth and used them to train the null hypothesis.
The remaining normal samples were used as part of the test batch, along with the samples from the two anomalous clusters.
This was repeated 10 times, with the performance averaged.

For the network data, all the normal web flows from the three files were combined, making nearly ten thousand normal web flows.
We randomly selected 20\% of these flows as ground-truth normal samples to train the null, 
and treated the remaining normal flows as part of the test batch, combined with either P2P or Zeus anomalous flows.
We separately experimented with P2P and Zeus flows.
There were roughly 5 \% of either P2P or Zeus flows in a given test batch.
Experiments for each scenario were averaged over 10 random train-test splits.
\subsection{Experimental results}

\begin{figure}
\centering
\begin{tabular}{ll}
\hspace{-4.5mm} \includegraphics[width=1.8in]{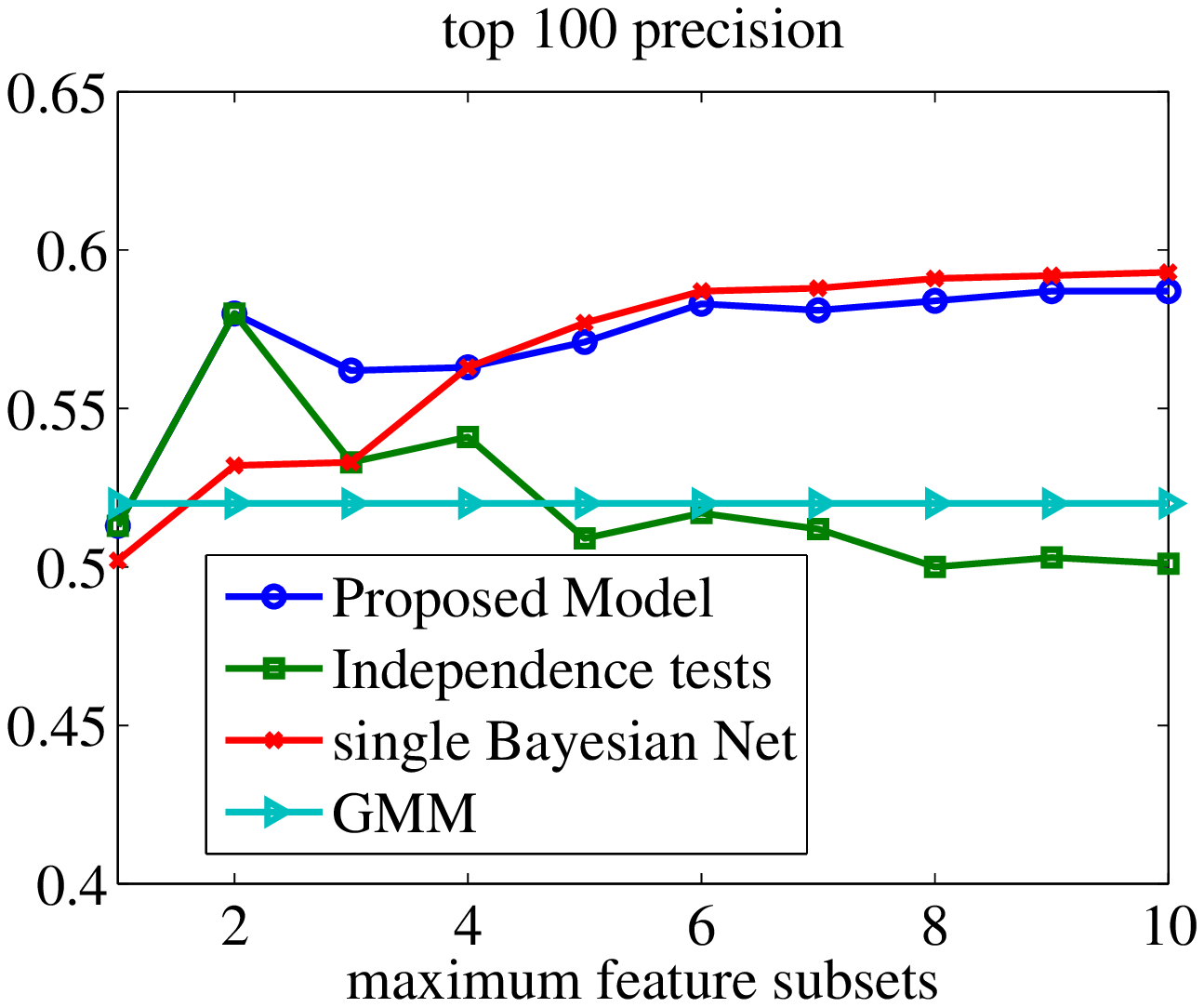}&
\hspace{-4.5mm} \includegraphics[width=1.8in]{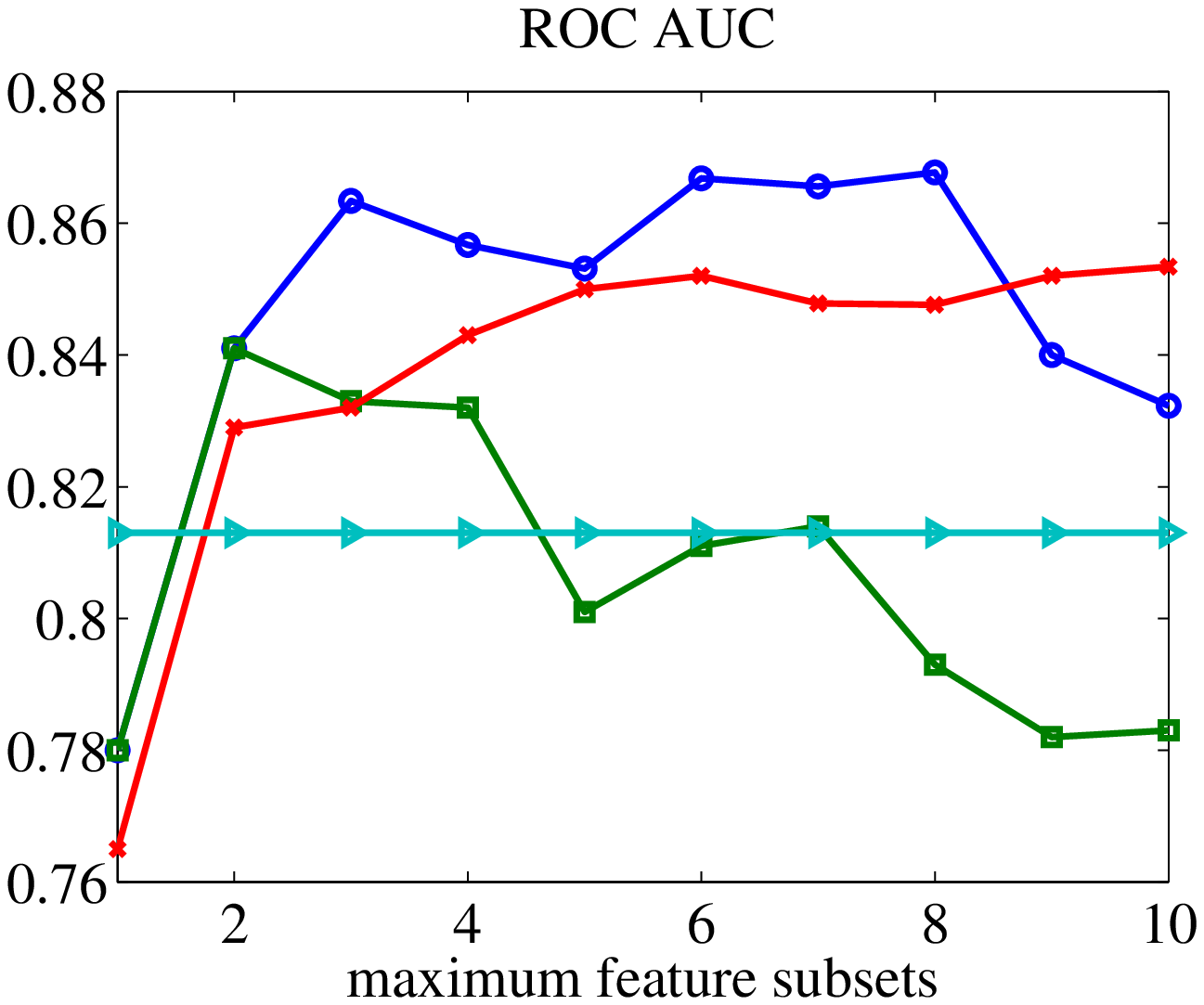} \\
\end{tabular} 
\caption{Synthetic data experiment: comparison of different schemes with 2 independent Gaussian based anomalous feature subsets in (separate) 1-dim subspace}
\label{syn}
\end{figure}

In Figure \ref{syn}, we show the performance on the synthetic data. 
Note that both the proposed scheme and \cite{Fatih} effectively capture groups of anomalies when the maximum feature subset order is two.
The first captured cluster (sample subset) consists of more than 95\% anomalous samples on average.
However, as the maximum feature subset order increases, the 
``independence tests" approach drops significantly in performance.
This is because too many (assumed to be independent) pairwise tests create many redundant features that are all used to evaluate cluster anomalousness; use of these redundant features de-emphasizes, within the score function, the important (low-order) feature subset.
Also, we see an early advantage of using cluster-specific DTs, compared to the single Bayesian Net approach.
It appears that if an anomalous process is strictly generated from a low order subspace and normal in other feature dimensions (as is the
case in this experiment) our cluster-specific DT approach
outperforms a single Bayesian Net approach.

\begin{figure}
\centering
\begin{tabular}{ll}
\hspace{-4.5mm} \includegraphics[width=1.8in]{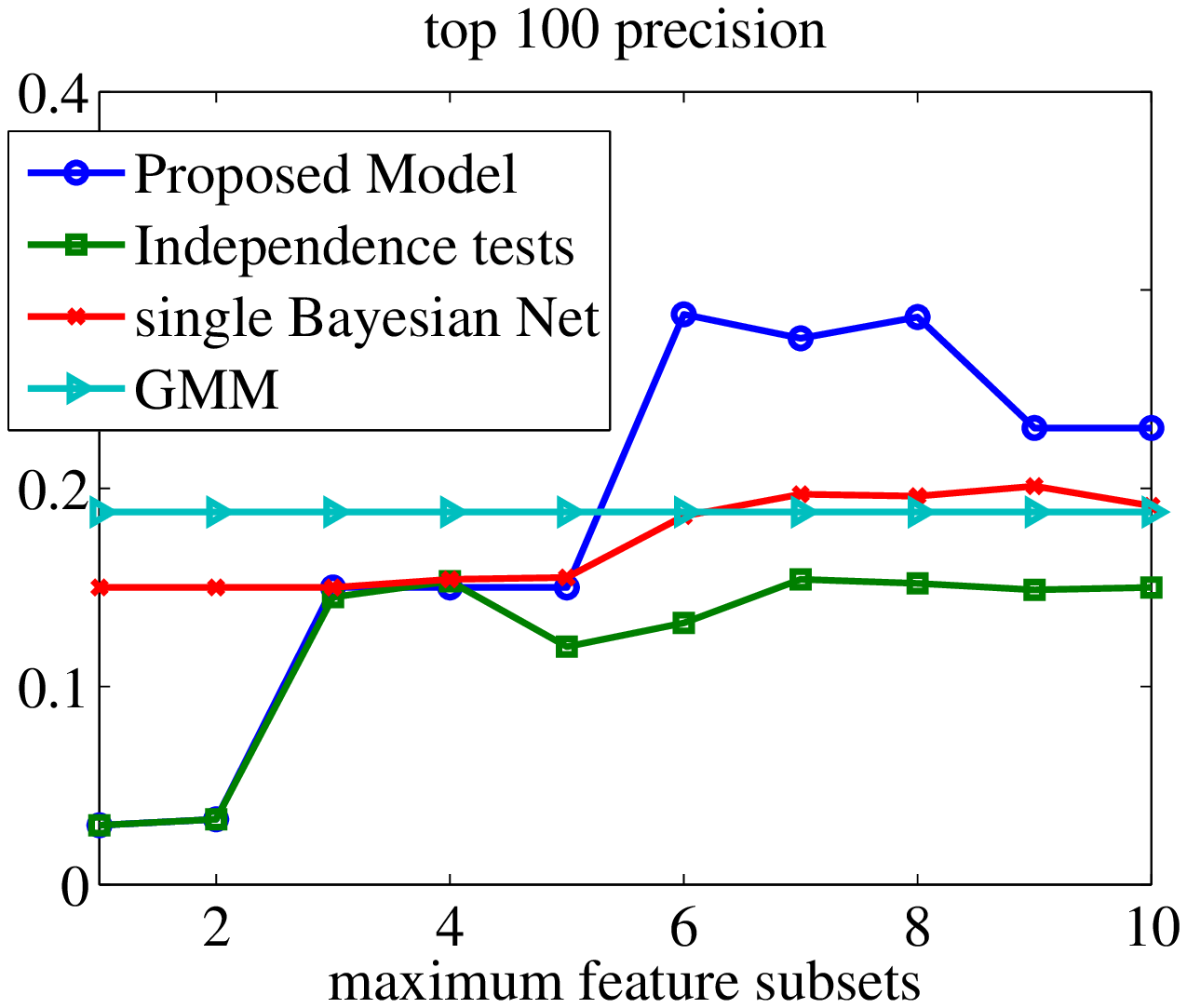}&
\hspace{-4.5mm} \includegraphics[width=1.8in]{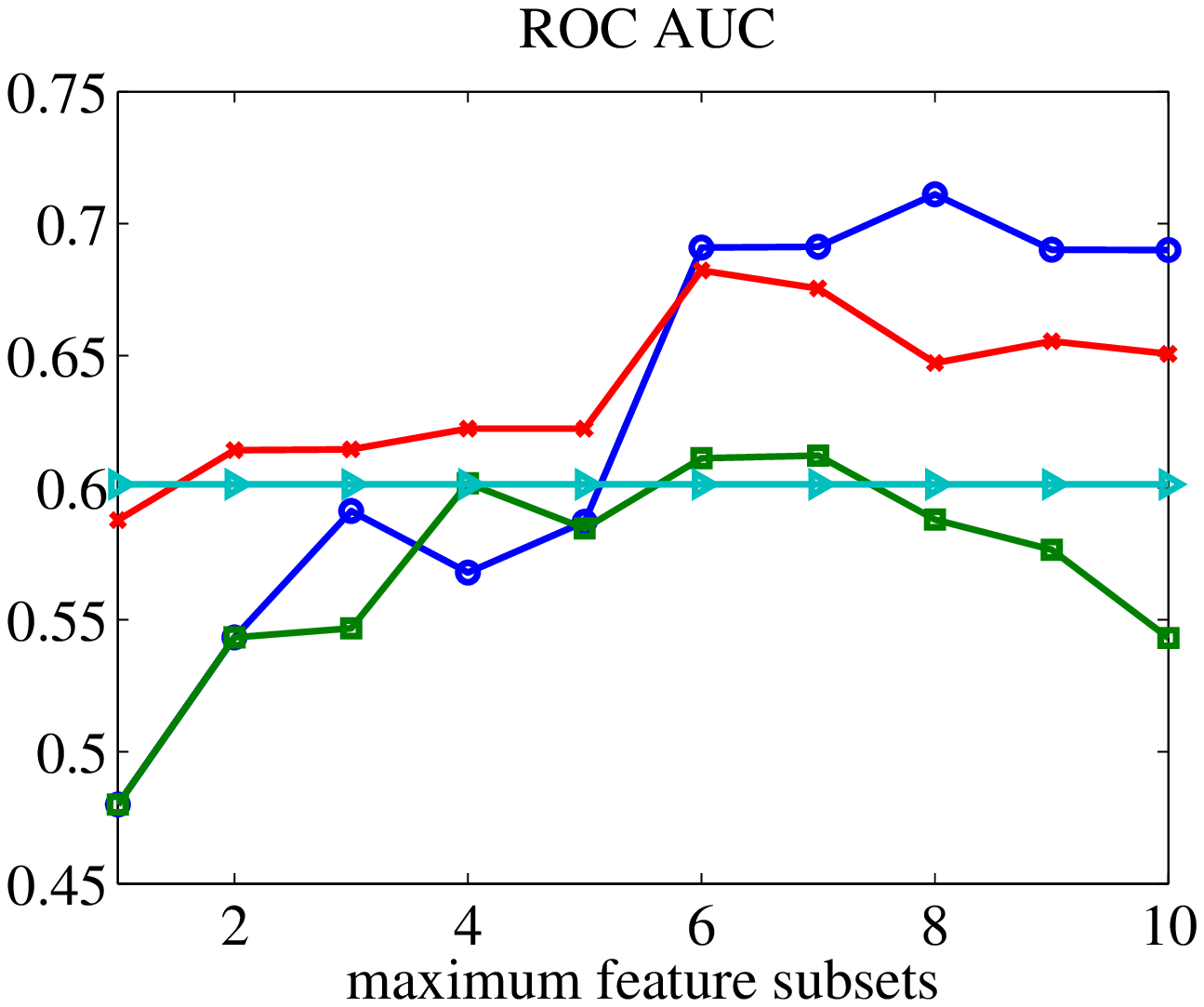} \\
\end{tabular}
(a) P2P 
\begin{tabular}{ll}
\hspace{-4.5mm} \includegraphics[width=1.8in]{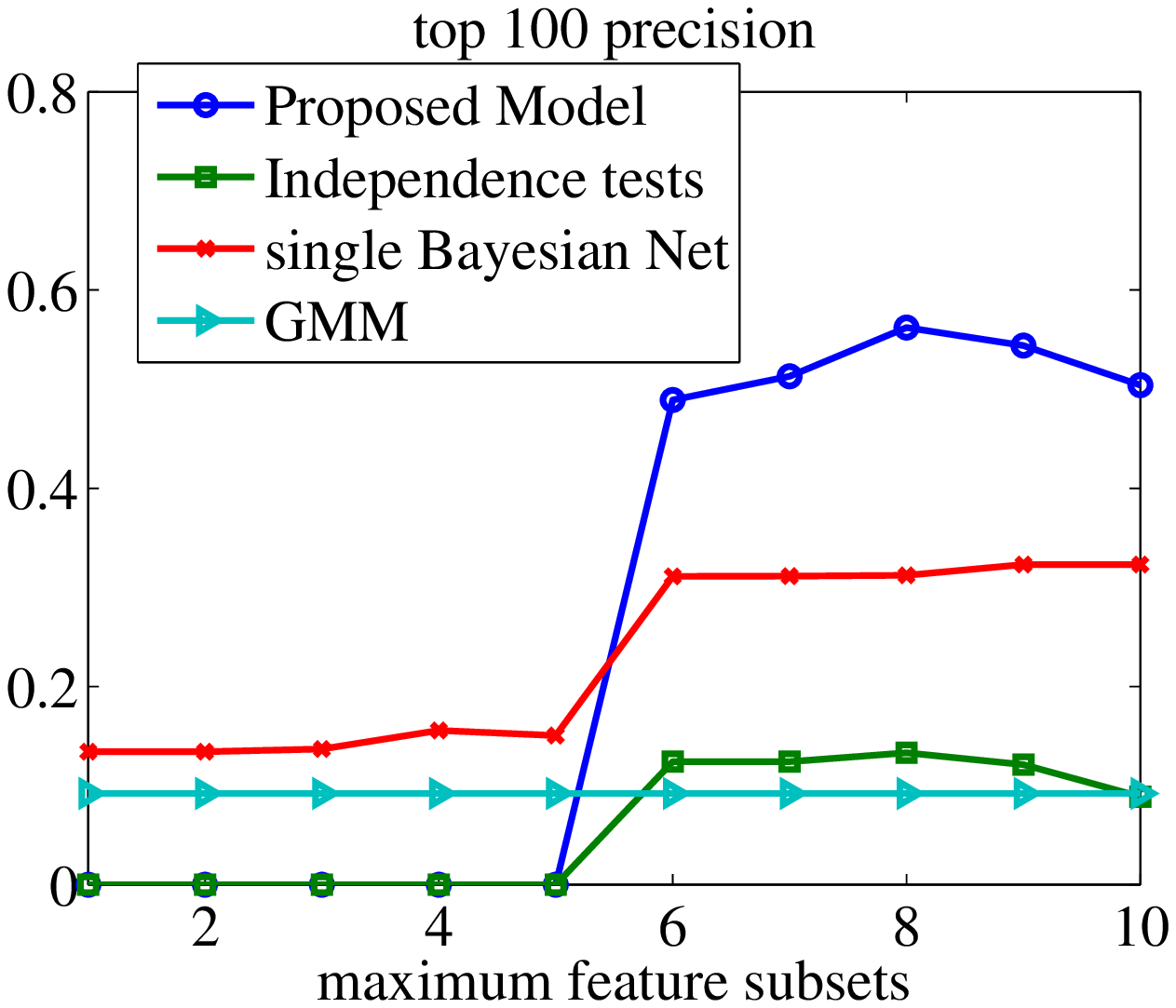}&
\hspace{-4.5mm} \includegraphics[width=1.8in]{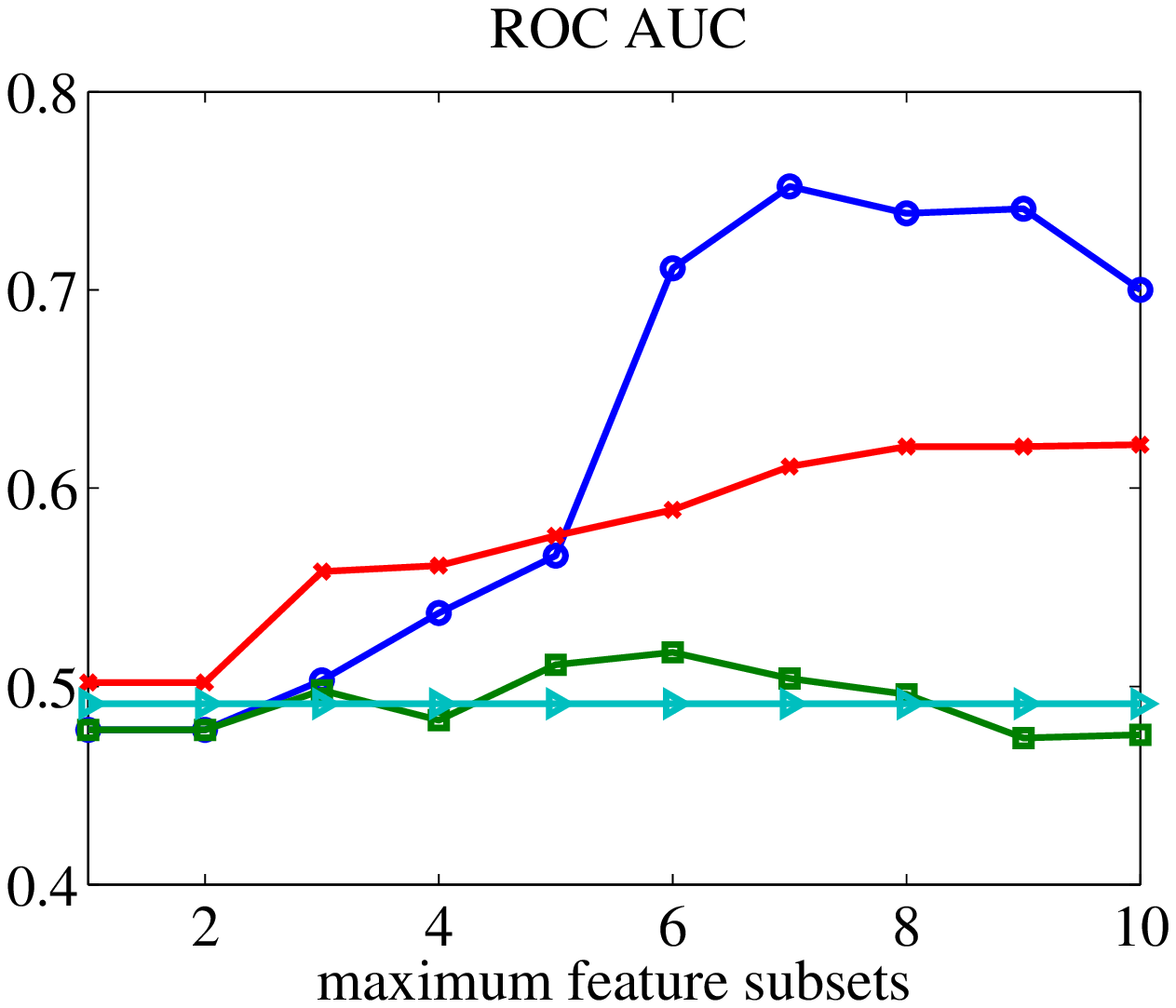} \\
\end{tabular}
(b) Zeus
\caption{Network traffic data experiment: comparison of different schemes with P2P or Zeus anomalies}
\label{exp}
\end{figure}

In Figure \ref{exp} a), we show the performance for normal-P2P discrimination. 
Compared to \cite{Fatih}, which degrades in performance as more and more tests are included, we see superior performance for the proposed method.
There is a large batch of anomalous samples captured at maximum order 6 by the proposed method, but both \cite{CMU} and \cite{Fatih} did not capture this group effectively, 
as seen in the top 100 precision figure. 
Also, both of these methods are outperformed by the GMM baseline method.
In Figure \ref{exp} b), we show the performance for normal-Zeus discrimination. 
Again, at maximum feature subset order 6 the proposed method captures a large portion of the anomalous flows --
more than 50 Zeus flows were captured out of the first 100 flows detected by the proposed method.
\cite{Fatih} performs poorly in this experiment, and again we observed that as the number of tests increase, 
the independence assumption degrades the detection performance.
The single Bayesian Net approach in \cite{CMU} also performs relatively poorly on this dataset.

\section{Extensions and Future Work}
In this work, we 
used the Bonferroni corrected score function to directly evaluate cluster candidates.  Alternatively, we could try to
evaluate empirical p-values for this decision statistic, by applying our detection strategy to (many) bootstrap test batches drawn from the null distribution.
It would be interesting to see whether such an approach gives comparable (or even better) detection accuracy than
use of the Bonferroni corrected score by itself.  Such an approach could also be used to determine whether any detected clusters
are truly statistically significant.
In this work we showed detection accuracy as a function of the maximum feature subset size for a cluster. 
As the maximum feature subset size continues to increase, we observed that false positives also increase in the first detected cluster,
and the objective in (\ref{bonf_corr}) tends to favor the maximum feature dimension over use of fewer dimensions.
In future, we should propose and investigate criteria for choosing this maximum feature subset size.
\section{Conclusion}
In this work, we proposed a GAD scheme to identify anomalous sample and feature subsets, accounting for dependencies between the
features in a given subset.
The proposed model outperforms previous works that assume statistical tests are independent under the null.
We demonstrated the effectiveness of our proposed system on both synthetic and real world data, with the latter
drawn from 
the network intrusion detection domain, aiming to discriminate between normal and P2P/Zeus traffic.
Our future work includes empirical p-value assessment and automatic determination of the maximum feature subset size of a cluster. 
\bibliographystyle{abbrv}
\bibliography{ref}

\begin{thebibliography}{10}

\bibitem{salt}
Z.~B. Celik, J.~Raghuram, G.~Kesidis, and D.~J. Miller.
\newblock Salting public traces with attack traffic to test flow classifiers.
\newblock In {\em Proc. USENIX CSET}, 2011.

\bibitem{Chow}
C.~Chow and C.~Liu.
\newblock Approximating discrete probability distributions with dependence
  trees.
\newblock {\em IEEE Transactions on Information Theory}, 14(3):462--467, 1968.

\bibitem{DAS}
K.~Das, J.~Schneider, and D.~B. Neill.
\newblock Anomaly pattern detection in categorical datasets.
\newblock In {\em Proc. 14th ACM SIGKDD International Conference on Knowledge
  Discovery and Data Mining}, pages 169--176, 2008.

\bibitem{Eskin2000anomaly}
E.~Eskin.
\newblock Anomaly detection over noisy data using learned probability
  distributions.
\newblock In {\em Proc. 17th International Conference on Machine Learning},
  pages 255--262. Morgan Kaufmann Publishers Inc., 2000.

\bibitem{Fatih}
F.~Kocak, D.~J. Miller, and G.~Kesidis.
\newblock Detecting anomalous latent classes in a batch of network traffic
  flows.
\newblock In {\em Proc. IEEE Conf. on Information Sciences and Systems (CISS)},
  pages 1--6, 2014.

\bibitem{LBNL}
{LBNL/ICSI Enterprise Tracing Project}.
\newblock \url{http://www.icir.org/enterprise-tracing}.

\bibitem{Li2009}
W.~Li, M.~Canini, A.~W. Moore, and R.~Bolla.
\newblock Efficient application identification and the temporal and spatial
  stability of classification schema.
\newblock {\em Computer Networks (Elsevier)}, 53(6):790--809, 2009.

\bibitem{efficient}
W.~Li and A.~W. Moore.
\newblock A machine learning approach for efficient traffic classification.
\newblock In {\em Proc. IEEE Int'l Symposium on Modeling, Analysis, and
  Simulation of Computer and Telecommunication Systems (MASCOTS)}, pages
  310--317, 2007.

\bibitem{CMU}
E.~McFowland, S.~Speakman, and D.~B. Neill.
\newblock Fast generalized subset scan for anomalous pattern detection.
\newblock {\em The Journal of Machine Learning Research}, 14(1):1533--1561,
  2013.

\bibitem{provisional}
D.~J. Miller and G.~Kesidis.
\newblock Detecting clusters of anomalies in a data batch that manifest on
  low-dimensional feature subsets via dependence-tree based evaluation of joint
  statistical significance.
\newblock {\em Provisional United States Patent Filing}, 2014.

\bibitem{Miller}
D.~J. Miller, F.~Kocak, and G.~Kesidis.
\newblock Sequential anomaly detection in a batch with growing number of tests:
  Application to network intrusion detection.
\newblock In {\em Proc. IEEE Int'l Workshop on Machine Learning for Signal
  Processing (MLSP)}, pages 1--6, 2012.

\bibitem{Paxson}
R.~Sommer and V.~Paxson.
\newblock Outside the closed world: On using machine learning for network
  intrusion detection.
\newblock In {\em IEEE Symposium on Security and Privacy}, pages 305--316,
  2010.

\bibitem{Zeus2}
{VRT Labs - Zeus Trojan Analysis}.
\newblock \url{https://labs.snort.org/papers/zeus.html}.

\bibitem{Hierarchical}
L.~Xiong, B.~P{\'o}czos, J.~G. Schneider, A.~Connolly, and J.~VanderPlas.
\newblock Hierarchical probabilistic models for group anomaly detection.
\newblock In {\em Proc. International Conference on Artificial Intelligence and
  Statistics}, pages 789--797, 2011.

\bibitem{GLAD}
R.~Yu, X.~He, and Y.~Liu.
\newblock {GLAD}: group anomaly detection in social media analysis.
\newblock In {\em Proc. 20th ACM SIGKDD International Conference on Knowledge
  Discovery and Data Mining}, pages 372--381, 2014.

\bibitem{Zou2011}
G.~Zou, G.~Kesidis, and D.~J. Miller.
\newblock A flow classifier with tamper-resistant features and an evaluation of
  its portability to new domains.
\newblock {\em IEEE Journal on Selected Areas in Communications},
  29(7):1449--1460, 2011.

\end{thebibliography}
\end{document}